\documentclass[%
reprint,
superscriptaddress,
showpacs,preprintnumbers,
 amsmath,amssymb,
pra,
]{revtex4-1}

\usepackage{natbib}
\usepackage{epsfig}
\usepackage{color}
\usepackage{ulem}
\usepackage{cancel}
\usepackage[utf8]{inputenc}
\usepackage{mathtools}

\usepackage{graphicx}% Include figure files
\usepackage{dcolumn}% Align table columns on decimal point
\usepackage{bm}% bold math
\usepackage[mathlines]{lineno}% Enable numbering of text and display math
\begin{document}

%\preprint{APS/123-QED}

\title{Analytical description of optical vortex beams generated by\\discretized vortex producing lenses  }% Force line breaks with \\

\author{Gonzalo Rumi}
%  \altaffiliation[Also at ]{Physics Department, XYZ University.}%Lines break automatically or can be forced with \\
\author{Daniel Actis}%

\affiliation{%
Centro de Investigaciones Ópticas, CONICET-UNLP-CIC, P.O. Box 3, 1897 Gonnet, Argentina
}%

% \collaboration{MUSO Collaboration}%\noaffiliation

\author{Dafne Amaya}
%  \homepage{http://www.Second.institution.edu/~Charlie.Author}
\affiliation{%
Centro de Investigaciones Ópticas, CONICET-UNLP-CIC, P.O. Box 3, 1897 Gonnet, Argentina
}%
% \affiliation{
%  Third institution, the second for Charlie Author
% }%
\author{Jorge A. Gómez}
\affiliation{%
Grupo de Física Básica y Aplicada, Politécnico Colombiano Jaime Isaza Cadavid, Medellín, Colombia
}%

\author{Edgar Rueda}
\affiliation{%
Grupo de Óptica y Fotónica, Instituto de Física, Universidad de Antioquia U de A, Calle 70 No. 52-21, Medellín, Colombia
}%

\author{Alberto Lencina}
 \email{Corresponding author: alencina@faa.unicen.edu.ar}
\affiliation{%
Laboratorio de Análisis de Suelos, Facultad de Agronomía, Universidad Nacional del Centro de la Provincia de Buenos Aires, CONICET, P.O. Box 47, 7300, Azul, Argentina
}%

% \collaboration{CLEO Collaboration}%\noaffiliation

\date{\today}% It is always \today, today,

\begin{abstract}

Discretized vortex-producing lenses programmed on low performance spatial light modulators have been used for the generation of optical vortices. However, the description of these vortices has been supported only by numerical simulations. In this work, a general analytical treatment (any topological charge - any discretization levels) for the propagation of a Gaussian beam through a discretized vortex-producing lens  is presented. The resulting field could be  expressed as a sum of Kummer beams with different amplitudes and topological charges focalized at different planes, whose characteristics of formation can be modified by tuning the parameters of the setup. Likewise, vortex lines are analyzed to understand the mechanism of formation of observed topological charges, which appear in specific planes. Conservation of the topological charge is demonstrated. Theoretical predictions are supported by experiments. 
% The understanding of the field generation and the vortex lines behavior in this kind of phase plates can be of great importance for compact vortex in-line applications. 

% \begin{description}
% Secondary publications and information retrieval purposes.
% \item[PACS numbers]
% 42.25.Fx,42.30.Kq.
% May be entered using the \verb+\pacs{#1}+ command.
% You may use the \texttt{description} environment to structure your abstract;
% use the optional argument of the \verb+\item+ command to give the category of each item. 
% \end{description}
\end{abstract}

\pacs{42.25.Fx,42.30.Kq.}% PACS, the Physics and Astronomy
\maketitle

%\tableofcontents

\section{\label{sec:Intro}Introduction}

Optical vortices have been widely studied due to their intrinsic characteristic of having optical orbital angular momentum \cite{Allen1992}. This particularity has allowed the use of this type of laser beams in a wide variety of fields, such as image processing \cite{Crabtree2004,Davis2000}, microscopy \cite{Furhapter2005}, optical tweezers \cite{Curtis2002,Arias2013}, optical communications \cite{Wang2016}, optical metrology \cite{Anzolin2009}, integrated optics \cite{Pu2015}, nanophotonics \cite{Toyoda2012}, among others.

Several methods have been proposed for optical vortex generation. Among them, the spiral phase plate (SPP) \cite{Lee2004,Furlan2009} is widely used, mainly due to its simplicity and high-energy efficiency. More recently, SPPs have been combined with radial dependence structures such as Fractal zone plates, Daman zone plates, and Fresnel lenses, allowing the creation of multi foci systems and the relaxation of phase modulation SPP conditions for the creation of high quality optical vortices   \cite{Tao2006,Yu2012,Rueda2013}. 

%{\color{red}(ER: yo eliminaría esto) In particular, Rueda \textit{et al.} \cite{Rueda2013} compared optical vortices generated by the diffraction of a fundamental $TEM00$ laser beam on a (SPP) with transmittance $\phi_{CON}(\theta)= e^{i\ell\theta}$ with the results on a vortex-producing lens (VPL) - SPP plus a Fresnel lens- with transmittance $\phi_{CON} (r, \theta)= e^{i\ell \theta - \frac{kr^2}{2 f_{FR}}}$, both displayed on a liquid-crystal spatial light modulator (LC-SLM)\cite{Rueda2013}. In these expressions $\ell$ is the vortex topological charge, $(r, \theta)$ are the polar coordinates, $k$ is the wave number, and $f_{FR}$ is the Fresnel lens focal distance.} 

In practical implementations, the ideal condition of a continuous phase modulation is usually not available due to technological limitations of spatial light modulators (SLMs) \cite{Moreno2004} and, as a consequence, the experimental implementation %of the vortex-creating systems 
requires the discretization of the mask. In the case of a discretized SPP (DSPP), it has been demonstrated theoretically and experimentally that, to achieve a high quality optical vortex, it is necessary a SLM with phase modulation capacity higher than $\frac{5\pi}{3}$, which is equivalent to have DSPPs of at least 6 phase levels ($N\geq 6$) \cite{Zhang2010,Guo2006}. In contrast, Rueda \textit{et al.} \cite{Rueda2013} demonstrated that using discretized vortex producing lenses (DVPLs) it is possible to generate high quality optical vortices of arbitrary topological charge using fewer phase levels. In fact, it was shown that by optimizing the experimental setup it is possible to generate high quality optical vortices with only two phase levels (phase modulation capacity of $\pi$) \cite{Londono2015}, meaning that low-modulation SLMs can be used.

Due to the angular periodicity of the DSPP, Guo \textit{et al.} \cite{Guo2006} expressed its phase as a Fourier series, interpreting therefore, the discrete mask as a linear combination of ideal continuous SPPs of different topological charge, whose associated coefficients depend on the beam \textit{principal} topological charge $\ell$ and the number  of phase levels $N$. In this sense, the field at the observation plane can be understood as a superposition of optical vortices of different topological charges converging to the same plane \cite{Guo2006}. In the case of the DVPL, the geometry of the phase mask includes a quadratic radial variation (Fresnel lens) that generates a multi-foci system whose diffraction properties will depend not only on the Fresnel lens parameters, but also on the $N$ discretization levels. 

In this work, we present a general analytical treatment (any topological charge - any discretization level) for the propagation of a Gaussian beam through a DVPL. In the procedure, the DVPL phase mask is decomposed as a combination of continuous SPPs. However, unlike the DSPP case, the coefficients of the series include an additional quadratic-phase term that explains why optical vortices of certain topological charges are observed at different planes along the optical axis. A complete analysis of the field propagation is developed in order to define the exact location, topological charge and energy contribution of each component. Results allow to describe the intrinsic mechanism associated to the generation of the multi-foci system. A  study of the mechanism associated to the formation of the optical vortices is performed by studying the vortex-lines and the conservation of the topological charge. Likewise, the formalism  allows to identify  relevant parameters of the setup to conveniently manipulate the diffraction conditions and, therefore, the energy structure along the optical axis. All theoretical predictions are supported by experimental results. These results not only provide insight on the basic physical mechanisms involved in the generation of optical vortices, but can also be of useful in engineering applications such as: optical angular momentum channels and compact vortex in-line metrological applications.

\section{DVPL Fourier expansion}
\label{sec: DVPL Fourier}

A vortex producing lens phase mask is expressed as
\begin{equation}
\Phi_{CON} (\rho, \phi)= \exp\left(i\ell \phi - i\frac{k\rho^2}{2 f_{FR}} \right),
\label{confase}
\end{equation}

\noindent where $(\rho, \phi)$ are polar coordinates, $ \ell$ is the topological charge (from now on it will be called \textit{principal} topological charge), $k$ is the wave-number, and $f_{FR}$ is the Fresnel lens focal distance. If the continuous phase profile is discretized in $\ell N$ phase steps and each step has a constant phase increment of $ \Delta\phi= \frac{2\pi}{N}$, the discretized complex transmittance of $\Phi_{CON}(\rho,\phi)$ will correspond to a DVPL phase mask, and is mathematically expressed  as
\begin{equation}
\Phi(\rho,\phi)= \exp\left(i\Delta\phi \:\text{Floor} \left[\frac{1}{\Delta\phi}(\ell \phi - \frac{k\rho^2}{2 f_{FR}})\right] \right),
\label{fase}
\end{equation}

\noindent being $\textnormal{Floor}[x]$ the function which takes the nearest integer smaller than or equal to $x$. Since the transmittance of Eq.(\ref{fase}) is a periodic function of the azimuthal angle $\phi$ with a period of $2\pi$,  it can be expanded into a Fourier series,
\begin{equation}
\Phi(\rho, \phi)=\sum_{m=-\infty}^\infty t_m (\rho) \exp\left( i m \phi \right),
\label{expfase}
\end{equation}

\noindent where the coefficients $t_m(\rho)$ depend on the radial variable $\rho$ and are obtained from: 
\begin{equation}
t_m(\rho) = \frac{1}{2\pi}\int_{0}^{2\pi} \Phi(\rho,\phi) \exp\left(-i m\phi \right) \textnormal{d}\phi .
\label{FS coeff}
\end{equation}

\noindent Solving the integral of Eq.(\ref{FS coeff}) (see Appendix \ref{Append A}) gives the following coefficients:
\begin{equation}
t_m(\rho)=\begin{dcases*}
\begin{aligned}
&\exp\left(-i\frac{mk\rho^2}{\ell2 f_{FR}}\right) \\[.2cm] &\times\exp\left(-i\frac{\pi m}{\ell N}\right) \textnormal{sinc}\left(\frac{\pi m}{\ell N} \right)
\end{aligned}& , $\frac{m - \ell}{N \ell} = 0,\pm 1, ...$ \\[.3cm]
\;0 & , otherwise
        \end{dcases*}
\label{ds FS coeff5}
\end{equation}

\noindent with $\textnormal{sinc}(x) =\sin(x)/x$. Then, the phase mask of Eq.(\ref{fase}) can be interpreted as a superposition of SPPs with topological charges
\begin{equation}
m=\ell(1+jN), \quad \text{where } j = 0, \pm1, \pm2, ....
\end{equation}

\begin{figure}[!t]
\includegraphics[scale=1,trim={0 0 0 0},clip,angle=0]{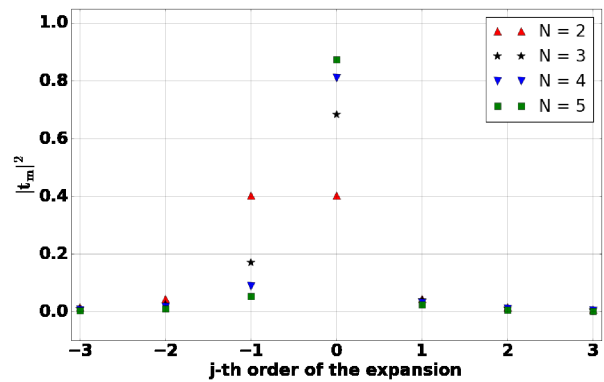}
\caption{Weight of terms in the expansion of Eq. (\ref{expfase}) as a function of the order $j$ for different $N$ levels, for any principal topological charge $\ell$.}
\label{orders}
\end{figure}

\noindent The weights $t_m$ have two phase factors: one of them is constant, whereas the other one has a radial quadratic dependency analogous to a Fresnel lens of focal distance $\frac{\ell}{m} f_{FR}$. The presence of this quadratic phase means that each SPP focuses at a different plane, depending on the number  of levels $N$ used and the  Fresnel lens focal distance. The number  of levels $N$ also plays a key role in the SPPs superposition, making the principal topological charge $\ell$ more dominant as $N$ increases. On the other hand, notice that the integer number $j$ refers to the different terms of the linear expansion without mentioning any specific topological charge. Thus, we call each term the $j$-th order of the expansion. Fig. \ref{orders} shows the weights for some orders of the expansion, for four different discretization levels. From the figure, some aspects can be emphasized: there is no weight symmetry with respect to the 0-th order, only for two-level discretization, orders -1 and 0, will have the same weight in the superposition and the same topological charge with different sign and, for four or more discretized levels the energy of higher orders are negligible.

% \begin{figure}[ht]
% \includegraphics[scale=0.22,trim={0 0 0 0},clip,angle=0]{js_eps.eps}
% \caption{Weight of terms in the expansion of Eq. (\ref{expfase}) for different $N$ levels, for any principal topological charge $\ell$.}
% \label{harmonics}
% \end{figure}

It is interesting to compare Eq. (\ref{ds FS coeff5}) with a very similar one that appears in reference \cite{Guo2006} for DSPPs (It corresponds to Eq. (\ref{ds FS coeff5}) but with $\rho = 0$). Because in \cite{Guo2006} the terms of the expansion are not accompanied by a quadratic phase factor, all the vortex of the linear combination are focused at the same $z$ plane, whereas for DVPLs they are focused at different planes, as it is shown in the next section.

\section{Field propagation after the DVPL} 
In this section the analytical expression for the field after crossing the DVPL is developed. The input plane is composed of the illumination beam $A(\rho)$ and the DVPL phase mask $\Phi(\rho,\phi)$, and can be written as
\begin{eqnarray}
U(\rho,\phi)=&& A(\rho)\bigg[\sum_{m=\ell+jN\ell}^{}\exp\left(-i\frac{mk\rho^2}{\ell2 f_{FR}}\right) \nonumber\\ 
&&\times \exp\left(-i\frac{\pi m}{\ell N}\right) \text{sinc}\left(\frac{m\pi}{N\ell}\right)\exp(im\phi)\bigg].
\label{input field}
\end{eqnarray}

\noindent This field is propagated a distance $f$ in free-space towards a  thin-lens of focal distance $f$. Then, it is propagated again an extra distance $z_0$. The complete analytical procedure is detailed in Appendix B. Here we report the result of the output field:
\begin{equation}
U(r,\theta;z_0)=\sum_m \exp\left(-i\frac{m\pi}{N\ell}\right) \text{sinc}\left(\frac{m\pi}{N\ell}\right)u_m(r,\theta;z_0),
\label{outputum}
\end{equation}

\noindent where $(r,\theta)$ are the output-plane coordinates and $u_m(r,\theta;z_0)$ has the form

% \begin{equation}
% \mathbf{u_n}(r,\theta)\propto e^{i n \theta} \mathcal{H}_m(\frac{kr}{f})[G(\rho) e^{\{i\frac{k}{2}(\frac{1}{f}-\frac{n}{\ell f_{SLM}}-\frac{z}{f^2})\rho^2\}}]
% \end{equation}

\begin{eqnarray}
&&u_m(r,\theta;z_0) = \frac{k \ i^{3m+1}}{f} \exp(ik[f + z_0])  \exp(im\theta) \nonumber\\ 
&&\times \mathcal{H}_m\bigg(\frac{kr}{f}\bigg) \left\{ A(\rho) \exp\bigg( \frac{ik\rho^2}{2}\bigg[\frac{1}{f} -\frac{m}{\ell f_{FR}} - \frac{z_0}{f^2}\bigg] \bigg) \right\}.
\label{propZcyluHtheory}
\end{eqnarray}

\begin{figure}[b]
\includegraphics[scale=1,trim={0 0 0 0},clip,angle=0]{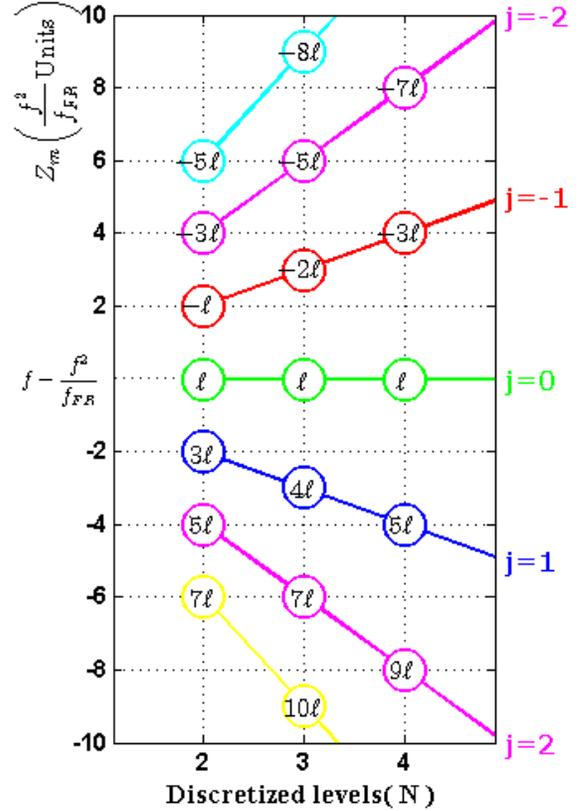}
\caption{
Position of the focused vortices as a function of the discretization level N for any principal topological charge $\ell$.}
\label{zm-f}
\end{figure}

Eq. (\ref{outputum}) is the resulting field at a distance $z_0$ from the physical thin lens, and it corresponds to a linear combination of the terms $u_m$, each one carrying a topological-charge $m$, and being proportional to the $m$--th order Hankel transform of a function $f(\rho)$ evaluated at $\kappa$, $\mathcal{H}_m(\kappa)\left\{f(\rho)\right\}$. Notice that for each term there exists a $z_0=z_m$ such that $-\frac{m}{\ell f_{FR}} + \frac{1}{f}-\frac{z_m}{f^2}=0$, and $\text{u}_m$ reduces to the Fourier transform of the original $m$--th term in the input field (Eq. (\ref{input field})). $z_m$ satisfies the expression
\begin{equation}
z_m=f-\frac{m}{\ell}\frac{f^2}{f_{FR}},
\label{focos}
\end{equation}

\noindent where $m=\ell, \ell \pm N\ell, \ell\pm 2 N \ell, \dots$. For the case of the principal topological charge, $m=\ell$, $z_m$ corresponds to the focus of the optical system formed by the physical lens and the Fresnel lens, while the other topological charges are focused at distances equal to multiples of $N\frac{f^2}{f_{FR}}$, with respect to the principal topological charge.
This can be easily seen by replacing the possible values of $m$ in Eq. (\ref{focos}) by the j-th orders, 
\begin{equation}
z_{m=\ell+\mathsf{j}N\ell} = \underbrace{f-\frac{f^2}{f_{FR}}}_{\text{focus of optical system}}-\underbrace{\mathsf{j}N\frac{f^2}{f_{FR}}}_{Additional foci}. 
\label{eq: z_m}
\end{equation}

Thus, for a given optical setup, it is possible to modify the distance between the focused vortex beams by changing the number $N$ of levels or by changing the focal distance of the lenses. Fig. \ref{zm-f} presents schematically this results. There, the position and charge of each  order is displayed as a function of the discretization level $N$. Note that the vertical axis is centered at the focus of the optical system $f-f^2/f_{FR}$ and the scale is in units of $f^2/f_{FR}$. From the figure, it is apparent that at the 0-th order, the principal vortex $\ell$ is obtained, irrespective the value of the discretization level $N$. As $N$ increases higher orders are more distant. This behavior, combined with the fact that higher orders carry less energy (see Fig. \ref{orders}), allows to understand why high quality vortex can be obtained with DVPL of lower $N$. Moreover, as the discretization increases the topological charge of the higher order increases, then increasing the dark disk contribution at the optical axis of the system. This situation, combined with the defocusing of each vortex away of the plane of its order, also contributes to improve the quality of the generated vortex. The later will be evident in the next section.

An interesting point to highlight is that because at each $z_m$ a different topological charge is focused, it seems that the topological charge is not conserved under propagation. However, as is discussed later, this is not the case.

\section{Gaussian beam input}

\begin{figure*}[ht]
\includegraphics[scale=.9]{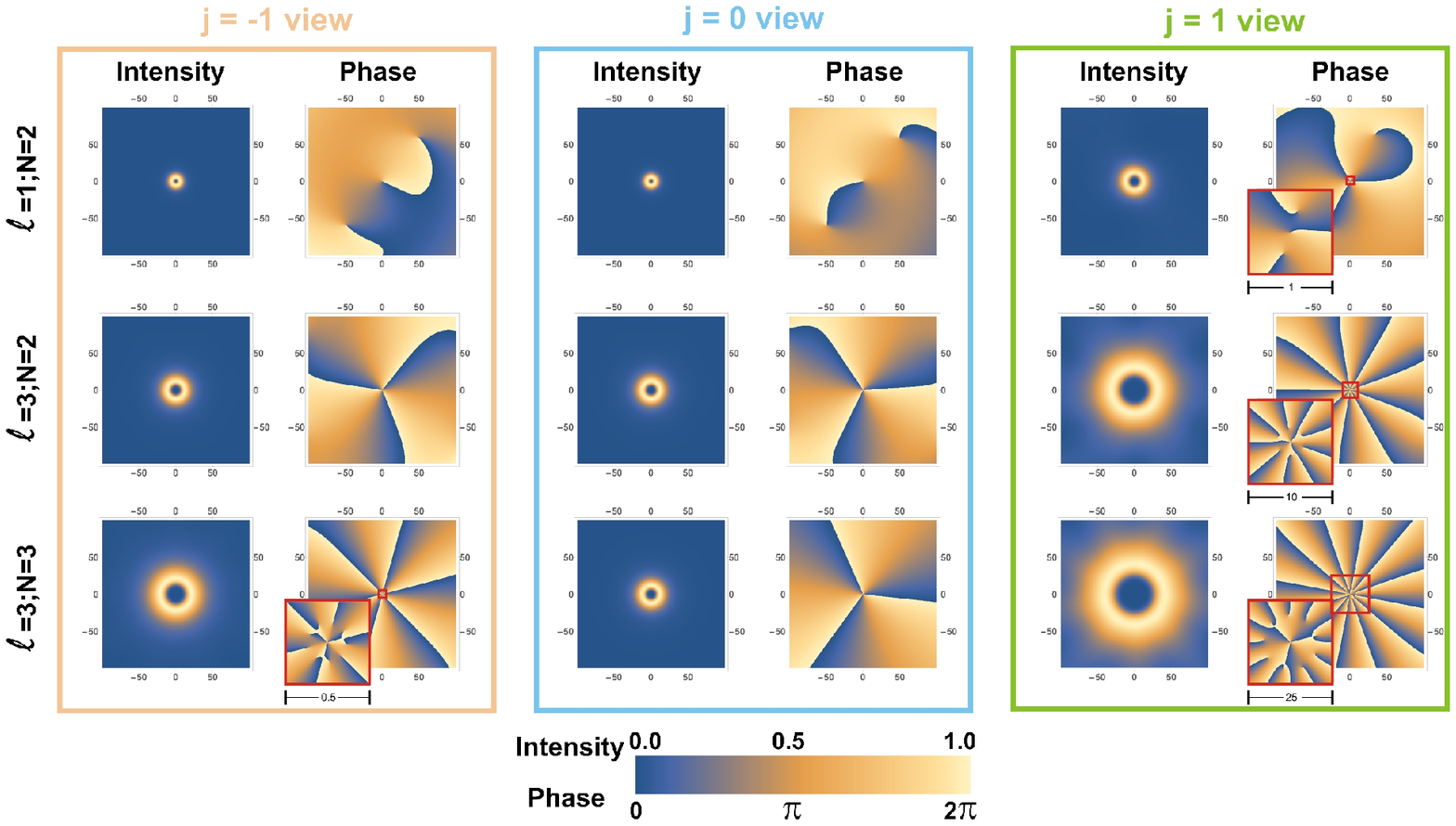}
\caption{Intensity and phase simulations for different DVPLs. For the simulations the following parameters were employed: $\lambda=532 \ \text{nm}$, $\omega_0=5 \ \text{mm}$, $f=20 \ \text{cm}$, $f_{FR}=1.6 \ \text{m}$. The results shown correspond to $N = 2$ and $3$, $\ell = 1$ and $3$, and $j = -1, 0$ and $1$.}
\label{fig:sim_int_pha}
\end{figure*}

For the special case of a gaussian beam, with beam waist $\omega_0$, and amplitude 
$$A(\rho) = \exp\big(-\frac{\rho^2}{\omega_0^2}\big),$$ 
by solving the Hankel transform \cite[Sec. 8.6, p. 29, Eq.(9)]{bateman1954tables}, the optical field at a distance $z_0$ is:

\begin{widetext}
\begin{eqnarray}
U(r,\theta;z_0) =&& \frac{k \sqrt{\pi}}{8f} \bigg(\frac{kr}{f}\bigg)\exp\big(ik(f+z_0)\big) \sum_m \frac{i^{3m+1} (-1)^{\frac{m-|m|}{2}}}{b_m^{3/2}} \exp(im\theta) \exp\bigg(-i\frac{m\pi}{N\ell}\bigg) \textnormal{sinc}\bigg(\frac{m\pi}{N\ell}\bigg)  \nonumber \\
&&\times \exp\bigg(-\frac{(kr/f)^2}{8b_m}\bigg) \bigg[ \textnormal{I}_{\frac{|m|-1}{2}}\bigg(\frac{(kr/f)^2}{8b_m}\bigg) - \textnormal{I}_{\frac{|m|+1}{2}}\bigg(\frac{(kr/f)^2}{8b_m}\bigg) \bigg],
\label{propaGauss}
\end{eqnarray}
\end{widetext}

\noindent where $\textnormal{I}_n(x)$ is the modified Bessel function, $m = \ell + j N\ell$ with $j=0,\pm1,\pm2,...$, and
$$b_m = \frac{1}{\omega_0^2} - \frac{ik}{2}\bigg(-\frac{m}{\ell f_{FR}} + \frac{1}{f} - \frac{z_0}{f^2}\bigg).$$

Each term of Eq. (\ref{propaGauss}) resembles a Kummer beam \cite{Anzolin2009}, with topological charge $m$, focused at $z_m$. Note that the Gaussian factor before the Bessel functions is the responsible to focus each term at its corresponding $z_m$. From this view, it is clear that the beam waist $\omega_0$ plays also an important role in defining the quality of the vortex at the 0-th order because it dominates the defocusing of the $\pm$1-th orders through $b_m$. It is interesting also to note that, although each term is focused at a corresponding $z_m$, the other terms are present and act as a background which could degrade the quality of the vortex by splitting the charge $m$ into $m$ unitary charges \cite{BEKSHAEV2004}. This point is addressed next.   

In Fig. \ref{fig:sim_int_pha} we present simulations of the resulting intensity and phase for the case of a DVPL illuminated by a gaussian beam, for different discretization levels, orders and principal topological charges. The parameters employed in the simulation are $\lambda = 532$ nm,  $\omega_0 = 5$ mm, $f = 20$  cm, and $f_{FR} = 1.6 $ m. % In Fig. \ref{fig:vortex_lines} we plot the vortex lines for the case of $N=3$ and $\ell=3$. 
The first thing to notice from Fig. \ref{fig:sim_int_pha} is that at a scale of the order of the doughnut-shaped intensity, the topological charge is as expected (see Fig. \ref{zm-f}). Yet, a closer look towards the optical axis shows that, except for the zero order (and the -1-th order for $N=2$), in all the other cases  there is a splitting of the topological charge $m$  into a bundle of vortices with lower topological charge. However, the principal charge $\ell$ prevails at the optical axis. This result suggest that the principal charge $\ell$ is robust against the background field corresponding to the remaining terms. This tells us that the topological charge of the principal vortex is conserved in the vicinity of the optical-axis (see the next section for a more detailed analysis). 
 
The case of $N=2$ is out of the previous analysis because the topological-charge at the optical axis reverses its sign for negative orders. This result could lead to think that topological-charge is not conserved under propagation. However, as it is well known, the dynamical inversion of the topological charge of an optical vortex occurs for noncanonical vortices in presence of astigmatic transformations \cite{Molina-Terriza2001,Bekshaev2002}. It was verified (not shown) that in between the planes of the -1-th and 0-th order there is a place where a Freund's critical foliation appears. Despite this intriguing phenomenon, the reversed topological charges appearing for negative orders are also robust against the background field corresponding to the remaining terms (not shown). Despite \textit{all} these striking phenomena, the overall topological charge is conserved (as expected) under propagation, as it is shown below.  
 
\begin{figure*}[!]
\includegraphics{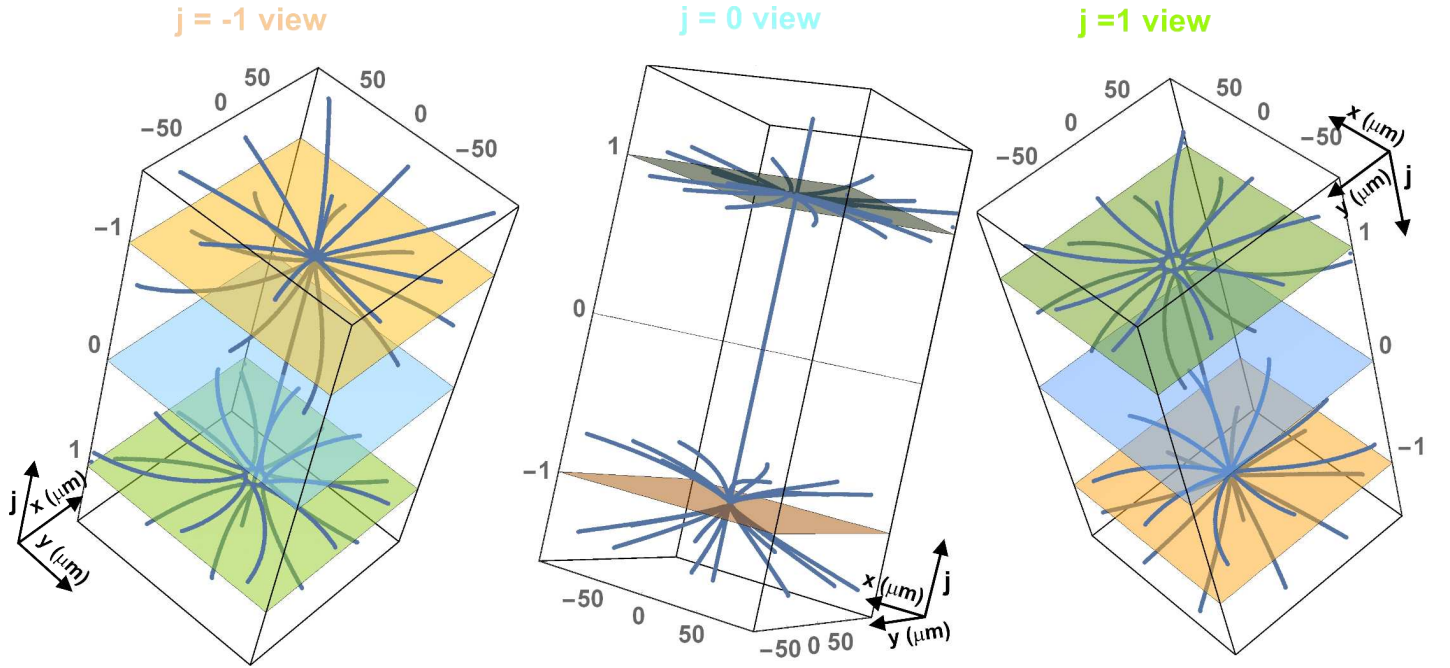}
\caption{Vortex lines obtained from a DVPL with $N=3$ and $\ell=3$. All figures represent the same result but different points of view. Each figure emphasizes an order plane ($j=-1, 0,1$). The employed parameters are the same as those of Fig. \ref{fig:sim_int_pha}.}
\label{fig:vortex_lines}
\end{figure*}

\section{Topological charge conservation}
\subsection{Vortex lines}

As it is well known, optical vortices are lines in space \cite{DENNIS2009,OHolleran2006}. In analogy with fluid mechanics \cite{saffman_1993}, the direction of these lines is described by the vorticity $\vec{\Omega}=\frac{1}{2}\vec{\nabla}\times\vec{\mathbf{j}}$, being $\vec{\mathbf{j}}$ the current associated with the field, i.e. the Poyting vector \cite{Berry2000,wang2005}. Since vorticity is defined by a curl, it is solenoidal, i.e. it has null divergence, which implies that vortex lines are closed curves. If a finite volume is considered, vortex lines close on themselves or the number of lines entering the volume are the same that leave it.

Figure \ref{fig:vortex_lines} shows the numerically obtained trajectories of vortex lines for a beam obtained with a DVPL with N=3 and $\ell=3$. The same parameters as in Fig. \ref{fig:sim_int_pha} are used. Three views of the same situation are depicted. In each view a plane is highlighted. Vortex lines enter and exit the region of interest, guaranteeing the conservation of topological charge from one plane to another. From the figure, it is apparent the presence of a principal vortex line at the optical axis and some vortex lines that concentrate around the $\pm 1$-th order planes. It is clear that all vortex lines that enter the volume also come out of it, in accordance with the definition of vorticity . For the $j=-1$ case, nine lines of topological charge $-1$ concentrate towards the principal vortex line of charge $\ell=3$. This gives rise to the $m=-6$ vortex observed at this plane, as predicted by Eq. (\ref{eq: z_m}) (see Fig. \ref{zm-f}) and shown in Fig. \ref{fig:sim_int_pha}. For $j=0$, only the principal vortex line is present, corresponding to an $\ell=3$ vortex.  In the case of $j=1$, again nine vortex lines concentrate close to the principal vortex line, but this time their topological charges are $1$, thus contributing to  the topological charge $m=12$ observed at this plane.

The previous analysis encompasses the concept of topological charge conservation from a geometrical point of view: if the only singularity introduced to the field is the principal topological charge then, because of the solenoidal property, the other must belong to closed curves so that the \textit{total} topological charge equals the principal one. This statement is analytically proven in the next subsection.

\subsection{Field at $r\rightarrow\infty$}
The topological charge  $\ell$ of a field is defined as the times of $2\pi$ its phase $\phi$ varies in a closed path, i.e. by calculating the Burgers vector as $\oint d\phi=2\pi\ell$. As it is observed from Figs. \ref{fig:sim_int_pha} and \ref{fig:vortex_lines} at $z_m$ the corresponding order dominates and the field strongly resembles  a vortex field of charge $m$. This fact could suggest that topological charge changes as the field propagates. However, as discussed in the previous section, the vortex lines are closed curves whose segments group together around the focus of an order. If this is what happens, the topological charge calculated with a path enclosing all loops must return the value of the topological charge imprinted in the DVPL. To ensure all loops are included the path must be taken at $r\rightarrow \infty$. By considering the first and second term of the asymptotic expansion of the Bessel functions \cite[Sec 10.40.1, p.255]{NISTHandbook2010}, the Eq. (\ref{propaGauss}) for $r\rightarrow\infty$ can be written as
\begin{eqnarray}
U(r_{\rightarrow\infty},\theta)=\frac{f}{k r^2} e^{i k (f+z_0)} \sum _m && i^{\left| m\right| +2 m+1} \left| m\right|\text{sinc} \left(\frac{\pi  m}{L N}\right)\nonumber\\ 
&&\times e^{-\frac{i \pi  m}{L N}}e^{i m \theta}.
\end{eqnarray}

\noindent Note that, despite a overall phase therm depending on $z_0$, the field at $r\rightarrow\infty$ does not change on propagation. This is a first evidence that any property determined from this expression is conserved as the field propagates in space. After some algebraic manipulation the field in the previous expression can be written as
\begin{widetext}
\begin{eqnarray}
U(r_{\rightarrow\infty},\theta)=\frac{m f}{k r^2}\frac{1}{\cos (\theta  \ell N)-\cos \left(\frac{\pi  \ell N}{2}\right)}&&\left\{
\cos \left[\frac{1}{2} \pi  \ell (N+1)\right] \cos \left[k (f+z_0)+\theta  \ell+\frac{1}{2} \pi  (2 \ell+1)-\frac{\pi }{N}\right]+\right.\nonumber\\[.3cm]
&&-\cos \left[\frac{\pi  \ell}{2}\right] \cos \left[k (f+z_0)+\theta  \ell (1-N)+\frac{1}{2} \pi  (2   \ell+1)-\frac{\pi }{N}\right]+\nonumber\\[.3cm]
&&+i \cos \left[\frac{1}{2} \pi  \ell (N+1)\right] \sin \left[k (f+z_0)+\theta  \ell+\frac{1}{2} \pi  (2 \ell+1)-\frac{\pi }{N}\right]+\nonumber\\[.3cm]
&&\left. -i\cos \left[\frac{\pi  \ell}{2}\right] \sin \left[k   (f+z_0)+\theta  \ell (1-N)+\frac{1}{2} \pi  (2 \ell+1)-\frac{\pi }{N}\right]\right\}.
\end{eqnarray}
\end{widetext}

\begin{figure*}[t]
\includegraphics{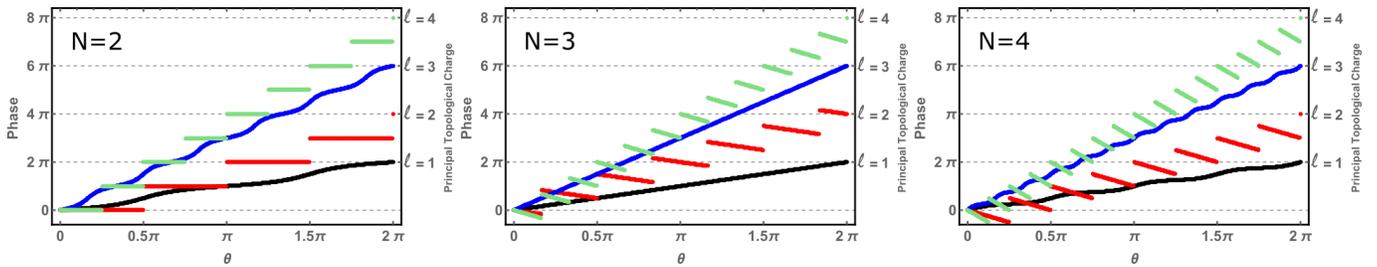}
\caption{Phase variation for $r\rightarrow \infty$ as a function of $\theta$ for different values of discretization level and principal topological charge. }
\label{fig:phase_r_inf}
\end{figure*}

\noindent From this expression the phase can be easily calculated. Fig. \ref{fig:phase_r_inf} shows the unwrapped phases for different values of discretization levels and principal charges. From these results, it is apparent that if the Burgers vector is calculated its magnitude coincides with the principal charge imprinted to the DVPL. These results prove that the topological charge is conserved under propagation irrespectively for any principal charge and discretization level considered.

\section{Experimental verification}

\begin{figure}[ht]
  \includegraphics[scale=0.3,trim={0 0 0 0},clip,angle=0]{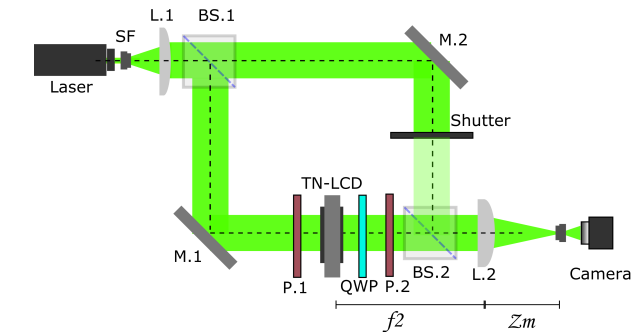}

\caption{Experimental setup to create and register optical vortex generated by a DVPL. The laser beam is collimated with an spatial filter SF, and a collimating lens L.1. A Mach-Zehnder is formed with beam-splitters BS.1 and BS.2, and mirrors M.1 and M.2. The DVPL is composed by a polarizer P.1, a quarter-wave plate QWP, a TN-LCD, and an analizer P.2. To observe the vortex a lens L.2, a microscope objective, and a camera are used. The shutter and reference arm are employed to recover the phase.}
  \label{ExpSetup}
\end{figure}

In order to experimentally validate the results obtained in the previous sections a DVPL  was built. The mounted experimental setup is presented in Fig. \ref{ExpSetup}. A laser of wavelength $\lambda = 532$ nm is filtered and collimated by a spatial filter SF and a lens L.1. This beam passes through a Mach–Zehnder interferometer formed by beam splitters BS.1, BS.2 and mirrors M.1 and M.2. One arm of the interferometer is used as a reference wave for phase retrieval, while on the other the DVPL was built up through a phase-mostly spatial light modulator, composed by a polarizer P.1, a quarter wave-plate $QWP$, a Holoeye LC2002 twisted-nematic liquid crystal display (TN-LCD), and an analyzer P.2. A lens L.2 is placed at a distance equal to the focal length $f_2 = 20$ cm from the TN-LCD. Finally, by using a 40x microscope objective, the intensity and the interference patterns of the complex field are registered with a CMOS camera (DCC1545M Thorlabs). Both, microscope objective and camera, could be moved longitudinally, in order to explore the $z_0$ dependence of the beam. A shutter is employed to block the reference beam when intensity images are recorded. The phase was recovered by using a five-step phase-shifting technique by encoding the phase delays directly in the LC-SLM \cite{Rastogi2000}.

% In order to validate the previous analysis, an experimental implementation for generating and detecting the field described by the expression \ref{propZcyluHtheory} was made it by using a transmission Twisted Nematic Spatial Light Modulator Based on Liquid Crystal Display (TN-LCD) in one of the arms of a Mach-Zendher interferometer. The TN-LCD have to be used in a phase-mostly configuration.This configuration involves complementing the TN-LCD with polarizers and one or two quarter-wave plates with their axes properly oriented [ref]. The experimental setup used is presented in Fig. (\ref{ExpSetup}).

\begin{figure*}[ht]
\includegraphics[scale=0.9]{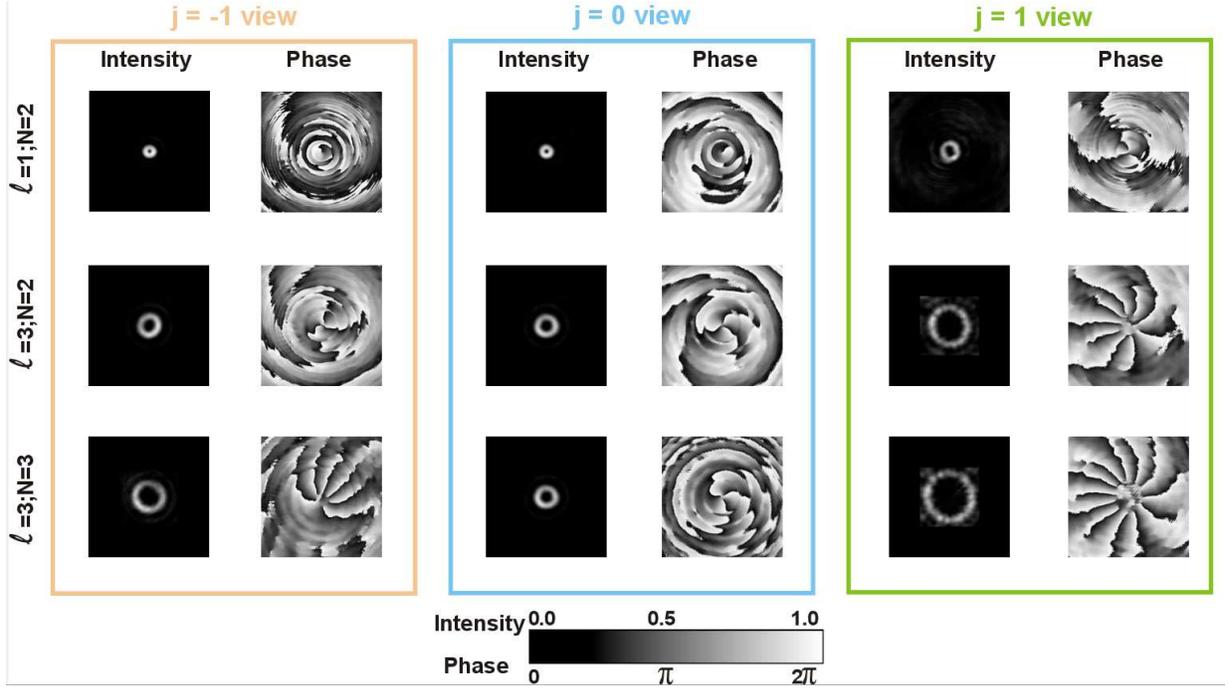}
\caption{Optical vortices obtained experimentally with DVPL implementations. For the experiments, the same parameters that in simulations were employed: $\lambda=532 \ \text{nm}$, $\omega_0=5 \ \text{mm}$, $f=20 \ \text{cm}$, $f_{FR}=1.6 \ \text{m}$. The results shown correspond to $N = 2$ and $3$, $\ell = 1$ and $3$, and $j = -1, 0$ and $1$. See Fig. \ref{fig:sim_int_pha} for comparison.}
\label{fig:exp_int_pha}
\end{figure*}

In order to build up the DVPLs, the Holoeye LC2002 TN-LCD was characterized in a phase-mostly configuration using the procedure developed by Amaya \textit{et al} \cite{amaya2017least}. A maximum phase-modulation close to $1.5 \pi$, with a $ 5 \%$ of coupled amplitude was obtained. This performance allows to implement DVPLs up to $N=3$. The same parameters of the simulations were employed for the experimental results: $\lambda=532 \ \text{nm}$, $\omega_0=5 \ \text{mm}$, $f=20 \ \text{cm}$, $f_{FR}=1.6 \ \text{m}$.  Results are shown in Fig. \ref{fig:exp_int_pha} for $N = 2$ and $3$, $\ell = 1$ and $3$, and $j = -1, 0$ and $1$. The experimental results are in excellent agreement with the predictions of the analytical expression of the field (Eq. (\ref{propaGauss})) and shown in Fig. \ref{fig:sim_int_pha}. For example, at the 0-th order, central column (j=0 view), the vortices obtained were registered at the focus of the optical system $f-(f^2/f_{FR})$ and it can be observed that their topological charges coincide with the the ones programmed in the DVPL independently of the number of the discretization levels $N$  and the value of the topological charge $\ell$. 

\section{Conclusions}

The beams resulting of the system proposed by Rueda \textit{et al} \cite{Rueda2013}, i.e. by using DVPLs, can be considered as a superposition of Kummer beams of different topological charges, each of which is focused at a different plane. The additional quadratic term in Eq. (\ref{ds FS coeff5}) generates a longitudinal separation of the focal plane of each Kummer. Simple expressions relating the relevant parameters of the system, the position of the foci and the appearing topological charges are derived. This allows to understand the presence of a high-quality vortex at $z_{\ell}$ reported by Rueda \textit{et al}: the most important term of Eq. (\ref{propaGauss}) is focused at that plane, while all other components are out of focus (their contribution to the intensity and the total phase of the beam are negligible at that plane). The presence of the additional foci and the topological charge of the phase singularity present at each of them was verified experimentally. This complete understanding of  the system helps us see that the role of the physical lens in the separation of the different components of the beam is irrelevant (although it ensures they be equidistant), and the quality of the system could be improved by omitting it. An interesting point is that, although this kind of system generates optical vortices of different topological charges in different planes along the propagation axis, it was possible to demonstrate that topological charge is a conserved quantity. This work also highlights the importance of the beams resulting from this system, given that they posses a complex topological structure. Finally, the understanding of vortex beam generation could be of importance in different applications, such as optical angular momentum channels and compact vortex in-line metrological applications.

%\renewcommand\thesection{APPENDIX \Alph{section}:}\setcounter{section}{0}
%\renewcommand{\thesection}{\Alph{section}}\setcounter{section}{0}
%\numberwithin{equation}{section}

\appendix
\section{\label{Append A}  Detailed calculation of the DVPL Fourier expansion}
% A VPL phase mask is expressed as
% \begin{equation}
% \Phi_{CON} (\rho, \phi)= \exp\left(i\ell \phi - \frac{k\rho^2}{2 f_{FR}} \right),
% \label{confase_apend}
% \end{equation}

% \noindent where $(\rho, \phi)$ are polar coordinates, $ \ell$ is the topological-charge, $k$ is the wave-number, and $f_{FR}$ is the Fresnel lens focal distance.  Discretizing the phase in $\ell N$ phase steps with constant phase-height increments of $ \Delta\phi= \frac{2\pi}{N}$, the discretized complex transmittance of $\Phi_{CON}(\rho,\phi)$ will correspond to a DVPL phase mask, expressed mathematically as
% \begin{equation}
% \Phi(\rho,\phi)= \exp\left(\Delta\phi \:\text{Floor} \left[\frac{1}{\Delta\phi}(\ell \phi - \frac{k\rho^2}{2 f_{FR}})\right] \right),
% \label{fase_apend}
% \end{equation}

% \noindent being $\textnormal{Floor}[x]$ the function which takes the nearest integer smaller than or equal to $x$. Since the transmittance of Eq. (\ref{fase_apend}) is a periodic function of the azimuthal angle $\phi$ with a period of $2\pi$,  
As was stated in Sec. \ref{sec: DVPL Fourier} a DVPL can be expanded into a Fourier series,
\begin{equation}
\Phi(\rho, \phi)=\sum_{m=-\infty}^\infty t_m (\rho) \exp\left( i m \phi \right),
\label{expfase_apend}
\end{equation}

\noindent where the coefficients $t_m(\rho)$ depend on the radial variable $\rho$ and are given by: 
\begin{eqnarray}
t_m(\rho) &&= \frac{1}{2\pi}\int_{0}^{2\pi} \Phi(\rho,\phi) \exp\left(-i m\phi \right) \textnormal{d}\phi \nonumber\\
&&= \frac{1}{2\pi}\int_{0}^{2\pi}\exp\left(i\Delta\phi \textnormal{Floor} \left[\frac{1}{\Delta\phi}\left(\ell \phi - \frac{k\rho^2}{2 f_{FR}}\right)\right] \right) \nonumber\\ 
&&\ \times \exp\left(-im\phi \right)\textnormal{d}\phi.
\label{FS coeff_apend}
\end{eqnarray}

By making the change of variable $\hat{\phi}=\frac{1}{\Delta\phi}(\ell \phi - \frac{k\rho^2}{2 f_{FR}})$, replacing $\Delta\phi$ by  $\frac{2\pi}{N}$, and taking out the terms that do not depend on the integration variable, Eq.(\ref{FS coeff_apend}) can be rewritten as,
\begin{eqnarray}
t_m(&& \rho) = \exp\left(-i\frac{mk\rho^2}{\ell2 f_{FR}}\right) \nonumber \\ 
&& \times \bigg[\frac{1}{N\ell}\int_{0}^{N\ell}\exp\left(i\frac{2\pi}{N} \textnormal{Floor}(\hat{\phi})-i \frac{m2\pi\hat{\phi}}{\ell N} \right)\textnormal{d}\hat{\phi}\bigg].\,
\label{int theta_hat_apend}
\end{eqnarray}

Calling the term in brackets $c_m$, for $0 \leq p < \ell N$, $p \in \mathbb{N}$, $\textnormal{Floor}(\hat{\phi})$ takes constant values in the interval $p \le \hat{\phi} < p+1$, thus
\begin{equation}
c_m = \frac{1}{\ell N} \sum_{p=0}^{\ell N - 1} \exp\bigg(i\frac{2\pi}{N} p\bigg)  \int_{p}^{p + 1} \exp\bigg(-i\frac{2\pi m}{\ell N}\hat{\phi} \bigg) \textnormal{d} \hat{\phi}.
\label{dspp FS coeff3_apend}
\end{equation}

Defining $\kappa = \frac{2\pi m}{\ell N}$, each of the terms in the sum of Eq. (\ref{dspp FS coeff3_apend}) can be expressed in the form:
\begin{equation}
\int_{-\infty}^{\infty} \textnormal{rect}\bigg(\hat{\phi} - \frac{2p+1}{p} \bigg) \exp(-i\kappa\hat{\phi}) \textnormal{d} \hat{\phi},
\label{rect int_apend}
\end{equation}

\noindent which is the Fourier transform of the function $\textnormal{rect}(\hat{\phi} - \frac{2p+1}{p})$ describing a rectangle of unit height. The solution is equal to
\begin{equation}
\exp\bigg( -i  \frac{2p+1}{p}\kappa \bigg) \textnormal{sinc}(\kappa/2).
\label{rect FT_apend}
\end{equation}
 
Replacing this result in Eq. (\ref{dspp FS coeff3_apend}), taking into account that $\textnormal{sinc}(z) = \sin(z)/z$, and using the definition of $\kappa$, $c_m$ can be written as

\begin{eqnarray}
c_m =&& \frac{1}{\ell N} \exp\bigg(-i\frac{\pi m}{\ell N}\bigg) \textnormal{sinc}\bigg(\frac{\pi m}{\ell N}\bigg) \nonumber \\ 
&&\times \sum_{p=o}^{\ell N - 1} \exp\bigg(\frac{i2\pi p}{N}\big(1-\frac{m}{\ell}\big)\bigg).
\label{dspp FS coeff4_apend}
\end{eqnarray}

If $\frac{m-\ell}{\ell N}$ is an integer n, all the terms in the sum of Eq. (\ref{dspp FS coeff4_apend}) are equal to $\exp(2\pi pn) = 1$, and so the sum is equal to $\ell N$. In any other case the sum is null. %Knowing this we can finally write $c_m$ in a compact form:
Finally, the coefficients of the expansion can be written as in Eq. (\ref{ds FS coeff5})
% \begin{equation}
% c_m = \begin{dcases*}
% \exp\bigg(-i\frac{\pi m}{\ell N}\bigg) \textnormal{sinc}\bigg(\frac{\pi m}{\ell N}\bigg) & , if $\frac{m - \ell}{N \ell} = 0,\pm 1, ...$ \\
% 0 & , otherwise
%         \end{dcases*}
% \label{dspp FS coeff5_apend}
% \end{equation}

% \noindent By replacing Eq. (\ref{dspp FS coeff5_apend}) in Eq. (\ref{int theta_hat_apend}), the coefficients of the expansion can be written in the following form:
% \begin{equation}
% t_m(\rho)=\begin{dcases*}
% \begin{aligned}
% &\exp\left(-i\frac{mk\rho^2}{\ell2 f_{FR}}\right) \\[.2cm] &\times\exp\left(-i\frac{\pi m}{\ell N}\right) \textnormal{sinc}\left(\frac{\pi m}{\ell N} \right)
% \end{aligned}& , $\frac{m - \ell}{N \ell} = 0,\pm 1, ...$ \\[.3cm]
% \;0 & , Otherwise
%         \end{dcases*}
% \label{ds FS coeff5_Apend}
% \end{equation}

\section{\label{Append B} DIFFRACTED-BEAM SPATIAL PROPAGATION}

For an optical field $U(\xi,\eta)$ at an input plane, the corresponding field, in the Fresnel approximation, at a distance $f$ is
\begin{eqnarray}
U(u,v) =&& \frac{\exp(ikf)}{i\lambda f} \exp\bigg(\frac{ik}{2f} (u^2 + v^2) \bigg) \nonumber \\ 
&&\times \int_{-\infty}^\infty \int_{-\infty}^\infty \bigg[U(\xi,\eta) \exp\bigg(\frac{ik}{2f} (\xi^2 + \eta^2) \bigg) \bigg] \nonumber \\
&&\times \exp\bigg(\frac{-ik}{f} [\xi u + \eta v]\bigg) \textnormal{d}\xi \ \textnormal{d}\eta,
\label{propL}
\end{eqnarray}

\noindent where $(u,v)$ are the plane-coordinates at distance $f$. The field $U(u,v)$ is then refracted by a physical thin-lens with phase $\exp\big(\frac{-ik}{2f} (u^2 + v^2) \big)$, and propagates a distance $z_0$. The corresponding beam is given by the expression
\begin{eqnarray}
U&&(x,y) = \frac{\exp(ikz_0)}{i\lambda z_0} \exp\bigg(\frac{ik}{2z_0} (x^2 + y^2) \bigg) \nonumber \\ 
&&\times \int_{-\infty}^\infty \int_{-\infty}^\infty \bigg[U(u,v)\exp\bigg(\frac{ik}{2} (u^2 + v^2) \bigg(\frac{1}{z_0}-\frac{1}{f}\bigg) \bigg)  \bigg]  \nonumber \\
&&\times \exp\bigg(\frac{-ik}{z_0} [u x + v y]\bigg) \textnormal{d}u \ \textnormal{d}v,
\label{propZ}
\end{eqnarray}

\noindent where $(x,y)$ are the coordinates of the observation plane, at a distance $f + z_0$ from the input one. Reorganizing terms in Eq. (\ref{propZ}) and integrating over coordinates $u$ and $v$ we obtain:
% \begin{eqnarray}
% U&&(x,y) = -\frac{\exp(ik[f + z_0])}{\lambda^2 f z_0}  \exp\bigg(\frac{ik}{2z_0} (x^2 + y^2) \bigg) \nonumber \\ 
% &&\times \int_{-\infty}^\infty \int_{-\infty}^\infty \bigg[U(\xi,\eta) \exp\bigg(\frac{ik}{2f} (\xi^2 + \eta^2) \bigg) \bigg] \nonumber \\ 
% &&\times \bigg\{\int_{-\infty}^\infty \int_{-\infty}^\infty \exp\bigg(\frac{ik}{2z_0} (u^2 + v^2) \bigg) \nonumber \\ 
% &&\times \exp\bigg(-ik \bigg[u\bigg(\frac{x}{z_0} + \frac{\xi}{f}\bigg) + v\bigg(\frac{y}{z_0} + \frac{\eta}{f}\bigg)  \bigg] \bigg)  \nonumber \\
% &&\times \textnormal{d}u \ \textnormal{d}v \bigg\}  \textnormal{d}\xi \ \textnormal{d}\eta,   
% \label{propZ2}
% \end{eqnarray}

\begin{eqnarray}
U&&(x,y) = \frac{\exp(ik[f + z_0])}{i \lambda f} \nonumber \\ 
&&\times \int_{-\infty}^\infty \int_{-\infty}^\infty \bigg[U(\xi,\eta) \exp\bigg(\frac{ik}{2f} \big(1 - \frac{z_0}{f}\big) (\xi^2 + \eta^2) \bigg) \bigg] \nonumber \\ 
&&\times \exp\bigg(-\frac{ik}{f} [x\xi + y\eta] \bigg)  \textnormal{d}\xi \ \textnormal{d}\eta.
\label{propZ3}
\end{eqnarray}

% Eq. (\ref{propZ3}) corresponds to the field at the observation plane, and for $z_0 = f$ it is the Fourier Transform of the input plane. 
\noindent By writing now Eq. (\ref{propZ3}) in cylindrical coordinates, $x=r\cos \theta$, $y=r\sin \theta$, $\xi=\rho \cos \phi$ and $\eta=\rho \sin \phi$, and using Eq. (\ref{input field}) as the input optical-field, it has
% \begin{eqnarray}
% U(r,\theta)&& = \frac{\exp(ik[f + z_0])}{i \lambda f}  \nonumber \\ 
% && \times \int_{0}^{2\pi}\int_{0}^\infty \bigg[U(\rho,\phi) \exp\bigg(\frac{ik}{2f} \big(1 - \frac{z_0}{f}\big) \rho^2 \bigg) \bigg] \nonumber \\ 
% &&\times \exp\bigg(-\frac{ik\rho r}{f} \cos(\theta - \phi) \bigg) \rho \textnormal{d}\rho \ \textnormal{d}\phi,
% \label{propZcyl}
% \end{eqnarray}

% and use Eq. (\ref{input field}) as the input optical-field,
% \begin{eqnarray}
% U(\rho,\phi)=&& A(\rho)\bigg[\sum_{m=\ell+jN\ell}^{}\exp\left(-i\frac{mk\rho^2}{\ell2 f_{FR}}\right) \nonumber\\ 
% &&\times \exp\left(-i\frac{\pi m}{\ell N}\right) \text{sinc}\left(\frac{m\pi}{N\ell}\right)\exp(im\phi)\bigg]
% \label{input field-2}
% \end{eqnarray}

% \noindent we obtain:
\begin{equation}
U(r,\theta) = \sum_m \exp\bigg(-i\frac{\pi m}{\ell N}\bigg) \textnormal{sinc}\bigg(\frac{m\pi}{N\ell}\bigg) u_m(r,\theta),
\label{propZcyl2}
\end{equation}

\noindent where

\begin{eqnarray}
u_m(r,\theta)&& = \frac{\exp(ik[f + z_0])}{i \lambda f} \int_{0}^\infty \bigg[\int_{0}^{2\pi} \exp(im\phi) \nonumber \\
&&\times \exp\bigg(-\frac{ik\rho r}{f} \cos(\theta - \phi) \bigg) \textnormal{d}\phi \bigg]  \exp\bigg(-i\frac{mk\rho^2}{2\ell f_{FR}}\bigg) \nonumber \\ 
&&\times \exp\bigg(\frac{ik}{2f} \big(1 - \frac{z_0}{f}\big) \rho^2 \bigg)A(\rho)  \rho \textnormal{d}\rho.
\label{propZcylu}
\end{eqnarray}

\noindent The integral in square brackets in Eq. (\ref{propZcylu}) can be evaluated using the identity \citep{Weisstein2017}

\begin{equation}
J_m(b) = \frac{i^{-m}}{2\pi} \int_0^{2\pi} \exp(im\alpha)\exp(ib\cos\alpha) \textnormal{d}\alpha, \ m = 1,2,...
\end{equation}

Thus we can write:

\begin{eqnarray}
 u_m&&(r,\theta) = \frac{ 2\pi\exp(ik[f + z_0]) i^m}{i \lambda f} \int_{0}^\infty \exp(im\theta) \nonumber \\
&&\times J_m\bigg(\frac{-k\rho r}{f}\bigg) A(\rho) \exp\bigg(-i\frac{mk\rho^2}{2\ell f_{FR}}\bigg) \nonumber \\ 
&&\times \exp\bigg(\frac{ik}{2f} \big(1 - \frac{z_0}{f}\big) \rho^2 \bigg)  \rho \textnormal{d}\rho.
\label{propZcylu2}
\end{eqnarray}

This solution is valid for $m = \pm 1, \pm 2, ...$ due to the identity $J_{-m}(s) = (-1)^m J_m(s)$ \citep{Weisstein2017}. Further, using the identity \cite{NISTHandbook2010} $J_m(b e^{i\pi}) = e^{i\pi m} J_m(b)$, we can write $J_m\bigg(\frac{-k\rho r}{f}\bigg) = i^{2m}J_m\bigg(\frac{k\rho r}{f}\bigg)$, and thus Eq. (\ref{propZcylu2}) can be simplified to:
% \begin{eqnarray}
%  u_m&&(r,\theta) = \frac{ k \exp(ik[f + z_0]) i^{3m+1}}{ f} \int_{0}^\infty \exp(im\theta) \nonumber \\
% &&\times J_m\bigg(\frac{k\rho r}{f}\bigg) A(\rho) \exp\bigg(-i\frac{mk\rho^2}{2\ell f_{FR}}\bigg) \nonumber \\ 
% &&\times \exp\bigg(\frac{ik}{2f} \big(1 - \frac{z_0}{f}\big) \rho^2 \bigg)  \rho \textnormal{d}\rho.
% \label{propZcylu3}
% \end{eqnarray}

% Organizing the last expression,
% \begin{eqnarray}
% u_m(r,\theta)&& = \frac{k \ i^{3m+1}}{f} \exp(ik[f + z_0]) \exp(im\theta) \nonumber \\ 
% &&\times \int_{0}^\infty \rho \bigg[ A(\rho) \exp\bigg(-i\frac{mk\rho^2}{2\ell f_{FR}}\bigg) \nonumber \\ 
% &&\times \exp\bigg(\frac{ik}{2f} \bigg\{1 - \frac{z_0}{f}\bigg\} \rho^2 \bigg)\bigg] J_m\bigg(\frac{k r}{f}\rho\bigg)   \textnormal{d}\rho,
% \label{propZcyluH}
% \end{eqnarray}

\begin{eqnarray}
u_m(r,\theta)&& = \frac{k \ i^{3m+1}}{f} \exp(ik[f + z_0]) \exp(im\theta) \nonumber \\ 
&&\times \int_{0}^\infty \rho J_m\bigg(\frac{k r}{f}\rho\bigg) \bigg[ A(\rho)  \nonumber \\ 
&&\times \exp\bigg(\frac{ik\rho^2}{2} \bigg\{\frac{1}{f} -\frac{m}{\ell f_{FR}}- \frac{z_0}{f^2}\bigg\}  \bigg)\bigg]   \textnormal{d}\rho,
\label{propZcyluH}
\end{eqnarray}

\noindent which has the form of a Hankel transform $\mathcal{H}_m\big(\frac{kr}{f}\big) \left\{ f_m(\rho) \right\}$, leading to Eq. (\ref{propZcyluHtheory}).

\bibliography{referencias}

\end{document}